# Understanding the size effects on the electronic structure of ThO$_2$ nanoparticles.

Lucia Amidani,*[a,b] Tatiana V. Plakhova,[c] Anna Yu. Romanchuk,[c] Evgeny Gerber,[a,b,c] Stephan Weiss,[b] Anna Efimenko,[d] Christoph J. Sahle,[d] Sergei M. Butorin,[e] Stepan N. Kalmykov[c] and Kristina O. Kvashnina*[a,b]

Developing characterization techniques and analysis methods adapted to the investigation of nanoparticles (NPs) is of fundamental importance considering the role of these materials in many fields of research. The study of actinide based NPs, despite their environmental relevance, is still underdeveloped compared to that of NPs based on stable and lighter elements. We present here an investigation of ThO$_2$ NPs performed with High-Energy Resolution Fluorescence Detected (HERFD) X-ray Absorption Near-Edge Structure (XANES) and with *ab initio* XANES simulations. The first post-edge feature of Th L$_3$ edge HERFD XANES disappears in small NPs and simulations considering non-relaxed structural models reproduce the trends observed in experimental data. Inspection of the simulations from Th atoms in the core and on the surface of the NP indeed demonstrates that the the first post-edge feature is very sensitive to the lowering of the number of coordinating atoms and therefore to the more exposed Th atoms at the surface of the NP. The sensitivity of the L$_3$ edge HERFD XANES to low coordinated atoms at the surface stems from the hybridization of the d-Density of States (DOS) of Th with both O and Th neighboring atoms. This may be a common feature to other oxide systems that can be exploited to investigate surface interactions.

## Introduction

Nanotechnology and nanomaterials are dominating the stage of scientific research since decades. When the size of a material is reduced to the nanoscale, exotic phenomena due to quantum confinement can appear and unique modifications of the optical, electronic, and mechanical properties show up, which can find application in fields spanning optoelectronic, catalysis, and medicine.[1] To understand and control the structure and the properties of nanomaterials, it is fundamental to develop a toolkit to adequately characterize these systems.[2]
Characterizing nanomaterials is a real challenge that requires a combination of different techniques as well as the development of analysis methodologies adapted to the very small particle's size.[3–5] The surface properties are of particular importance and very hard to investigate. X-rays Photoemission Spectroscopy (XPS), methods associated with Transmission Electron Microscopy (TEM), and total electron yield X-ray Absorption Spectroscopy (XAS) allow for structural and electronic characterization of the surface. However, the need of ultra-high vacuum prevents in situ investigations, the characterization of the sample under more realistic conditions, and in the case of oxides can induce changes in the local atomic environment, surface morphology and partial reduction.[6] In these regards X-ray Absorption Near-Edge Structure (XANES) and Extended X-ray Absorption Fine Structure (EXAFS) in the hard X-ray regime are very powerful techniques since they probe the local geometry and the local electronic structure of selected atomic species without imposing severe constrains on the sample conditions.[7–9] At the surface, the local properties differ drastically from those of the bulk material. When the size is progressively reduced and the ratio of surface to core atoms increases, size effects on the local geometry and on the electronic structure become visible in XANES and EXAFS spectra. EXAFS and XANES have indeed been widely applied to nanosystems, especially in pioneering fields like metal clusters for catalysis.[10,11] The appearance of size effects in XANES does not follow a unique and general behaviour. XANES spectra of nanoparticles (NPs) often present a broadening of spectral features compared to their bulk analogue, but trends specific to the system under investigation can be also observed and each case needs an *ad hoc* analysis to be correctly interpreted. Size effects most often appear as small variations in the NP's spectrum compared to that of the bulk system. Direct interpretation can be achieved only to a small extent with fingerprint approaches and the support of modelling and simulations are necessary to extract as much information as possible about the respective system. The use of High-Energy

[a.] Rossendorf Beamline at ESRF – The European Synchrotron, CS40220, 38043 Grenoble Cedex 9, France.
[b.] Helmholtz Zentrum Dresden-Rossendorf (HZDR), Institute of Resource Ecology, PO Box 510119, 01314 Dresden, Germany.
[c.] Department of Chemistry, Lomonosov Moscow State University, Leninskie Gory 1/3, 119991, Moscow.
[d.] ESRF – The European Synchrotron, CS40220, 38043 Grenoble Cedex 9, France.
[e.] Molecular and Condensed Matter Physics, Department of Physics and Astronomy, Uppsala University, P.O. Box 516, Uppsala, Sweden.

Resolution Fluorescence Detected (HERFD) XANES instead of conventional XANES can be crucial when the investigation relies on small spectral differences.[12,13] When XANES is acquired by integrating a portion of the characteristic X-ray fluorescence with a bandwidth smaller than the core-hole lifetime broadening, XANES features are sharpened and the detection of small spectral variations is facilitated.[14] Some recent works illustrated how the use of HERFD XANES boosts the investigation of nanoscale materials to probe not only their structure but also the dynamics of charge carriers.[15–20]

In the past, EXAFS was the technique of choice to investigate the structure of nanomaterials, but recently the advantages of XANES over EXAFS when investigating nanosystems have been stressed.[7,21,22] XANES is extremely sensitive to the absorber's local structural, i.e., bond distances, bond angles and the overall symmetry,[23] and it is a direct probe of the local electronic structure of the selected species with high sensitivity to the chemical state and local charge transfers. The quality of XANES is less affected by structural and thermal disorder and it can be collected on more diluted systems. The complexity of XANES interpretation has been for years the main bottleneck to its more diffused use as a characterization technique. The last decades have seen a remarkable improvement of *ab initio* codes dedicated to XANES, to Density Functional Theory (DFT) and to Molecular Dynamics (MD) and computer resources have constantly increased. These progresses have boosted the capabilities to interpret XANES and recently *ab initio* XANES simulations were used to train a neural network capable of retrieving the 3D structure of metal nanoparticles from experimental XANES.[22]

The great potential of XANES applied to nanomaterials is well established in fields where nanotechnology is well developed and the controlled synthesis, the characterization and the theoretical modelling of nanomaterials are advanced.[21] The potential of this technique remains mostly unexplored in fields where the nanotechnology is at its infancy like that of actinide-based nanomaterials. NPs are of fundamental concern in nuclear material research: they have a primary role in the migration of radionuclides in contaminated sites and nanostructuring the nuclear fuel pellets can potentially improve their mechanical and thermal properties.[24,25] The interest in actinide-based NPs is increasing,[26–31] however the size and shape effects on the chemical and physical properties are poorly investigated compared to systems made of lighter and stable elements. The reasons being the safety issues of handling radioactive samples during experiments and the problem of modelling heavy atoms with a partially filled f-shell theoretically.[29,32]

The use of ultra-high vacuum techniques on actinide-based nanoparticles is not trivial and measurements at ambient pressure are preferable. Despite approaches to XANES interpretation combining DFT and MD with *ab initio* codes for XAS are not yet applicable to actinide-based NPs,[32] simpler approaches can give important insight into the structural and chemical properties of these systems. Moreover, HERFD XANES in the hard X-rays is particularly advantageous for actinide-based nanomaterials since the gain in resolution at the L₃ edge of actinides is huge and the penetration depth of X-rays allows measuring in air and using multiple kapton confinements.

In this work, we present for the first time the investigation of $ThO_2$ NPs of different sizes with Th $L_3$ edge HERFD XANES. We observed a size effect in the post-edge region of spectra corresponding to NPs with a diameter below 3 nm and we applied a systematic approach based on *ab initio* simulations to link the changes observed in the electronic structure with the structure and the local environment of atoms at the surface. We found that the observed effect is a signature of surface Th atoms which are particularly exposed and for which the number of coordinating anions and cations are severely reduced compared to those characterizing bulk $ThO_2$.

## Experimental methods

### Nanoparticle synthesis

Samples were prepared by a chemical precipitation technique with subsequent drying under different conditions.[33] The primary particles were obtained with the following procedures: (A) 0.1M aqueous solution of thorium nitrate pentahydrate ($Th(NO_3)_4 \times 5H_2O$) was added under continuous stirring to a 3M sodium hydroxide aqueous solution. The obtained precipitate was repeatedly washed with MilliQ water to achieve the neutral pH value and divided in two equal parts. The two parts were dried for 12 hours in air in a drying oven at 40°C and at 150°C, respectively. (B) Aqueous solutions of 1M $Th(NO_3)_4 \times 5H_2O$ and 3M ammonia were mixed under continuous stirring, the obtained precipitate was washed and divided similarly to sample (A). The two parts were annealed in a muffle furnace for 4 hours at 400°C and 800°C, respectively.

### Characterization

The phase composition of the obtained powders was determined by X-ray diffraction (XRD) measurements with a Bruker D8 Advance diffractometer using a Cu Kα source (wavelength 1.54 Å). High-Resolution Transmission Electron Microscopy (HRTEM) images were acquired with Jeol-2100F HRTEM operated at 200 kV. The particle size was estimated from both XRD data and HRTEM images. The Scherrer equation was used to determine the $ThO_2$ NPs size from the XRD peaks broadening. The diameters of more than 200 well-defined NPs from HRTEM images were used to estimate the average $ThO_2$ particle size. The full results are reported in Ref. [33] and the HRTEM and size estimation of the smallest NPs are reported in the electronic supplementary information (ESI).

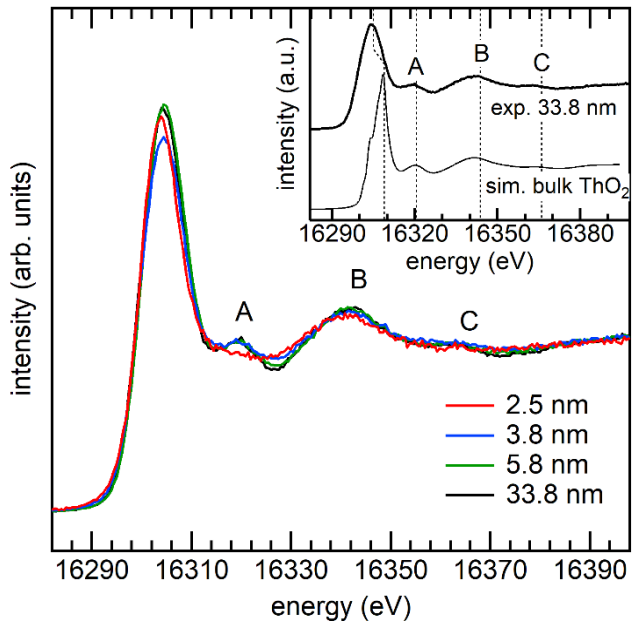

**Figure 2.** Main panel: experimental Th $L_3$ edge HERFD XANES on ThO$_2$ NPs of different sizes. Inset: HERFD XANES on 33.8 nm NPs and XANES simulation of bulk ThO$_2$ obtained with FDMNES. Vertical lines are added as guide to the eye to mark the main spectral features.

Th $L_3$ edge spectra on ThO$_2$ NPs were measured on the ROBL beamline at the ESRF.[34] The incident energy was selected with a Si(111) double crystal monochromator, the size of the beam at the sample was 400 μm horizontal times 150 μm vertical. The HERFD XANES were collected with an X-ray emission spectrometer[35] in Rowland geometry equipped with one Ge(880) spherically bent (0.5 m) crystal analyzer[36] to measure the maximum of the Th $L\alpha_1$ characteristic fluorescence line resulting in a total energy resolution of 2.9 eV. The Th $L_3$ edge HERFD XANES of ThO$_2$ bulk collected with improved resolution was measured on ID20 beamline at the ESRF.[37] The incident energy was selected with a cryogenically cooled Si(111) double crystal monochromator and a successive Si(311) channel-cut monocromator, the size of the beam at the sample was 20 μm horizontal times 10 μm vertical. The maximum of the Th $L\beta_5$ characteristic fluorescence line was selected with an X-ray emission spectrometer in Rowland geometry equipped with a Si(10 10 0) diced (1 m) crystal analyser. The overall resolution was 0.5 eV. Samples for HERFD XANES measurements were prepared as dried powders and wet pastes and sealed with single kapton confinement.

**Computational details**

Simulations of Th $L_3$ edge HERFD XANES were done with the FDMNES code.[38] The parameters of the simulation were tuned to obtain the best agreement for the HERFD XANES of bulk ThO$_2$ and then applied to simulate the HERFD XANES of structural models of ThO$_2$ NPs. Spin-orbit coupling and relativistic effects were included. The Fermi energy was estimated by a Self-Consistent Field (SCF) cycle including only the first oxygen coordination shell around the absorber, while the potential for the XANES calculation was built without SCF loop. The absorber was set to be excited and the density of states (DOS) projected on the absorber was also calculated. The error on the Fermi energy estimation was corrected by 3 eV for all simulations to have it in the middle of the energy gap. The convolution parameters *Gamma_hole* and *Gamma_max* were set to 1 eV and 15 eV, respectively, to obtain simulations comparable to HERFD XANES data on NPs, while *Gamma_max* was reduced to 10 eV to better match the spectrum of bulk ThO$_2$ acquired with higher resolution. The cluster radius, the method of calculation, and the use of SCF were carefully tested. We found that with a full multiple scattering (FMS) radius of 7 Å, selecting 98 atoms around the absorber, all the spectral features are well reproduced and increasing it further results only in minor variations of the spectral shape. Very similar results were found using the Green's function method and the Finite Difference Method (FDM) as well as with and without SCF. In particular, the post-edge region, which is the focus of the present work, was well reproduced with the Green's function method and without the need of SCF. We therefore used a FMS radius of 7 Å, no SCF and the Green's function method for all models of NPs considered. To investigate the effect of reducing the size of ThO$_2$ and of different NP shapes, we built three model NPs with tetrahedral, octahedral, and spherical shape cutting them from a chunk of ThO$_2$. We chose the NP size as close as possible to 2.5 nm, i.e., the average diameter for which we observe a size effect on the experimental data. Each model NP was simulated at once by using the keywords *all_conv* that calculates the spectrum of all inequivalent Th atoms present in the structure and their relative shifts of the Fermi level in order to appropriately calculate the weighted average.

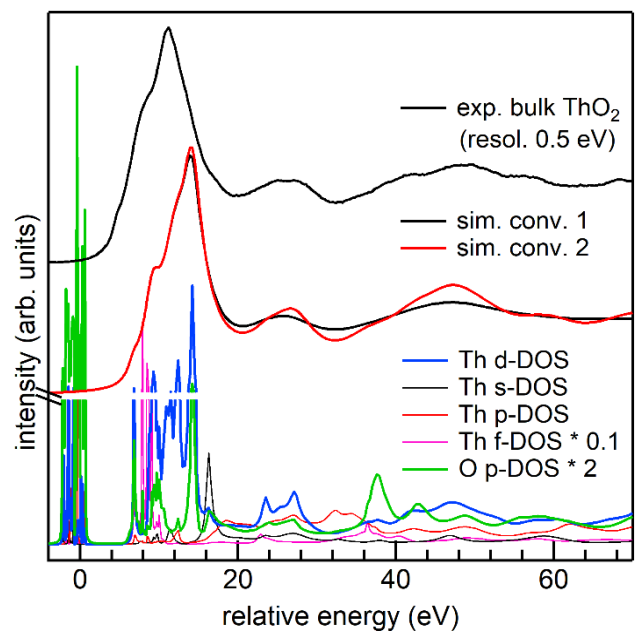

**Figure 1.** HERFD XANES collected with 0.5 eV resolution (top, black line) compared with two simulations of the $L_3$ edge XANES of Th in bulk ThO$_2$ obtained with different convolution parameters. The simulation in red is less convolved and the spectral features are in very good agreement with experimental data on top of the graph. On the bottom the partial DOSs of Th absorber and of the neighbouring O atoms are shown.

## Results and Discussion

According to XRD data,[33] crystalline NPs of 2.5±0.3, 3.8±0.4, 5.8±0.6 and 33.8±3.3 nm in diameter were obtained depending on the synthesis conditions. The average crystallite size of the first two samples was also determined from HRTEM images. Average diameters of 2.7±0.4 and 4.0±0.6 were found, which are consistent with the XRD results within the error. The XRD and HRTEM of the smallest NPs are reported in the ESI and all details on the synthesis and characterization of the samples can be found in ref. [33]. Figure 1 shows the experimental HERFD XANES data at Th $L_3$ edge of a series of such NPs. While numerous investigations reported the use of HERFD for U $L_3$ edge XANES,[13,39,40] its application to Th $L_3$ edge XANES represents a novelty. The XANES spectrum of the 33.8 nm NPs was found identical to that of bulk $ThO_2$ and will be used as reference. Inspection of the post-edge region along the series reveals a trend with decreasing size: the spectrum of 5.8 nm NPs is still identical to that of bulk $ThO_2$, when reducing the size to 3.8 nm the post-edge features A, B, and C are slightly broadened and feature A is less pronounced. Finally, for 2.5 nm NPs the broadening is more pronounced and feature A disappears. To understand the origin of the size effect observed in Figure 1, we focused on the smallest NPs and tried to reproduce the disappearance of feature A and the broadening of the rest of the post-edge using FDMNES. The simulation of bulk $ThO_2$ is shown together with the experimental spectrum of 33.8 nm NPs in the inset of Figure 1, where the simulation has been shifted to best match the post-edge features of the experimental spectrum. The agreement is very good, with the features A, B and C well reproduced in terms of energy position and relative intensity. The shape and the intensity of the main absorption peak, the white line (WL), are also in good agreement with the experimental data. We notice that the simulation predicts two shoulders on the rising edge of the WL that are not resolved in the data acquired on NPs, however they appear in HERFD XANES spectra acquired with higher resolution on bulk $ThO_2$ and shown in Figure 2. The simulation underestimates the separation between the WL and feature A, a common situation for codes dedicated to XANES. One possible reason is the strength of the 2p core-hole potential, not attracting enough the unoccupied 6d states of Th. We considered the agreement satisfactory for the scope of this work and avoided the use of special parameters to tune the strength of the screening. The same parameters used for bulk $ThO_2$ were applied to simulate the NPs. In particular, we chose the Green's function method because the use of the Finite Difference Method (FDM) was affecting mostly the WL at the expense of a much longer computational time.

The projected DOS obtained with FDMNES helps to understand the origin of the spectral features observed. $L_3$ edge XANES probes transitions from the 2p core state to the unoccupied DOS with d- and s- symmetry. In Figure 2, the HERFD XANES acquired with higher energy resolution (0.5 eV) on bulk $ThO_2$, the FDMNES simulations and the DOSs of Th and O are compared. Two simulations are shown, one matching the data on NPs (same as in Figure 1) and one with reduced convolution to match the experimental data acquired with higher resolution and shown in Figure 2. The agreement between the high-resolution data and the simulation is excellent: three shoulders on the rising absorption edge and substructures of feature A and B are well resolved in the data and well reproduced by the simulation. The shape of the simulated XANES follows the d-DOS of Th both at the WL and in the post-edge. The p-DOS of O contributes to some of the features of the WL and it presents many features in common with the Th d-DOS in the post-edge, indicating the presence of hybridization. We notice that the spectrum of bulk $ThO_2$ presented in Figure 2 differs from data

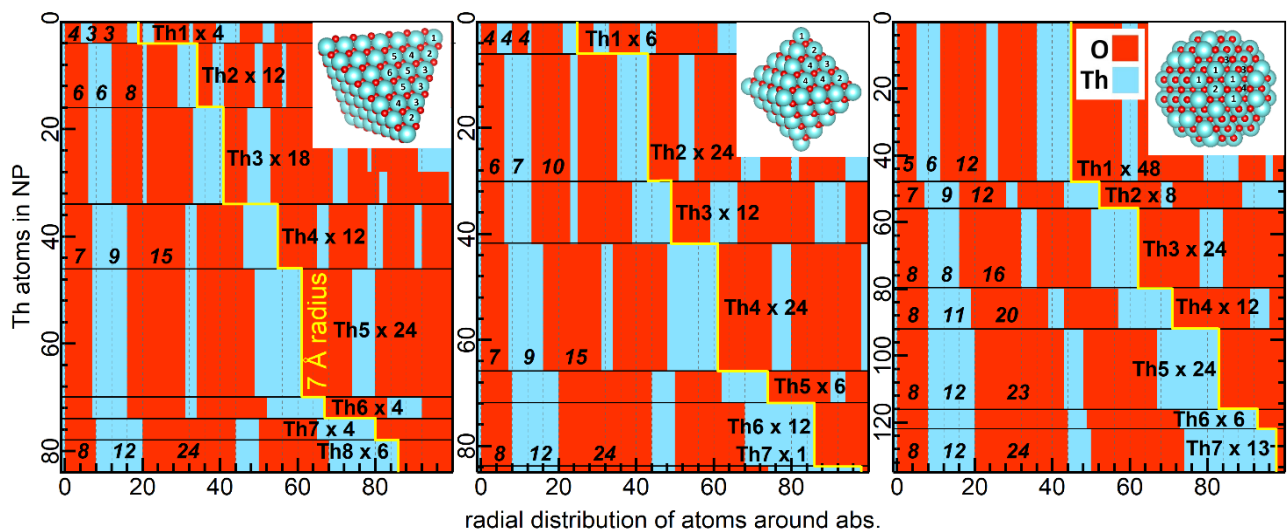

**Figure 3.** 2D images depicting the local environment of all Th atoms in the tetrahedral, octahedral, and spherical model NPs. The first element of each row is a Th atom of the NP, all the rest of the atoms in the NP are ordered according to increasing distance from the Th leading the row. Oxygen atoms are in red and Th atoms in light blue. The different coordination shells are easily recognized as well as the groups of equivalent Th atoms whose label Th# and multiplicity are indicated. The coordination numbers of the first three coordination shells are indicated. A yellow line delimits the atoms within a 7 Å radius from the Th leading the row.

presented in ref 41 despite it was obtained with better energy resolution.

The capability of the FDMNES code to precisely reproduce the spectral features observed in the experimental data for bulk $ThO_2$ justifies its use to investigate the origin of the size effect observed in the data. XANES is extremely sensitive to the local structure of the absorber, which is drastically different for Th atoms in bulk $ThO_2$ and Th at the surface of a NP. To understand at what extent the atoms at the surface can affect the XANES, we built model NPs from $ThO_2$ bulk and simulated with FDMNES the resulting XANES. We built three model NPs of different shapes and with size as close as possible to 2.5 nm. We considered a tetrahedral, an octahedral and a spherical nanoparticle.

The three models used in the calculations are shown in Figure 3. Together with the more general case of a spherical NP, we considered octahedral and tetrahedral models. The octahedral and the tetrahedral NPs expose only {111} facets, i.e., the most energetically favourable surface for many oxides with fluorite structure,[29,42,43] and are expected to be the most favourable shapes at small size.[44] The competition between these two shapes in case of small $CeO_2$ NPs has been investigated theoretically by Migani et al..[45] They found that below 3 nm the tetrahedral shape is even more favourable than the octahedral. $CeO_2$ NPs have the same crystal structure as $ThO_2$ and often serve as analogues of $ThO_2$ NPs. We cut the model $ThO_2$ NPs from a chunk of $ThO_2$ bulk without applying relaxation and charge compensation. We believe that this approach is indeed very useful to evaluate the effects of the shape and of the surface atoms on the XANES lineshape before introducing disorder. The resulting tetrahedral NP has 2.38 nm edges and is a $Th_{84}O_{172}$ cluster with 4 oxygen atoms in excess; the octahedral NP has the two opposite corners at 2.24 nm distance and is a $Th_{85}O_{160}$ cluster with 5 thorium atoms in excess. Finally, the spherical NP has a diameter of 2.1 nm and is a $Th_{135}O_{280}$ cluster with 10 oxygen atoms in excess.

Differently from $ThO_2$ bulk where all Th atoms are equivalent, in a NP different groups of non-equivalent Th cations are present due to the break of symmetry introduced by the surface. The shape and the size of the NP define a specific collection of non-equivalent Th atoms, each of which is characterized by a specific local environment. Figure 3 illustrates it schematically for the three model NPs we investigated: an image is built where each Th atom in the NP is in turn occupying the first element of the row. The row is then filled with the atoms composing the NP ordered by increasing distance from the Th leading the row. Only the first 100 atoms of each row are shown. This is enough to display all atoms within 7 Å distance, corresponding to the cluster cut-off considered in our calculations and indicated with a yellow line. Oxygen and thorium atoms are in red and light blue, respectively. This schematic view allows to distinguish at a glance the different layers of anions and cations surrounding the Th absorbers and referred to as coordination shells in the following. The scheme in Figure 3 allows also to easily group the Th with the same local environment within the 7 Å distance. We can distinguish several non-equivalent Th in each model: 8 in the tetrahedral, 7 in the octahedral and 7 in the spherical NP.

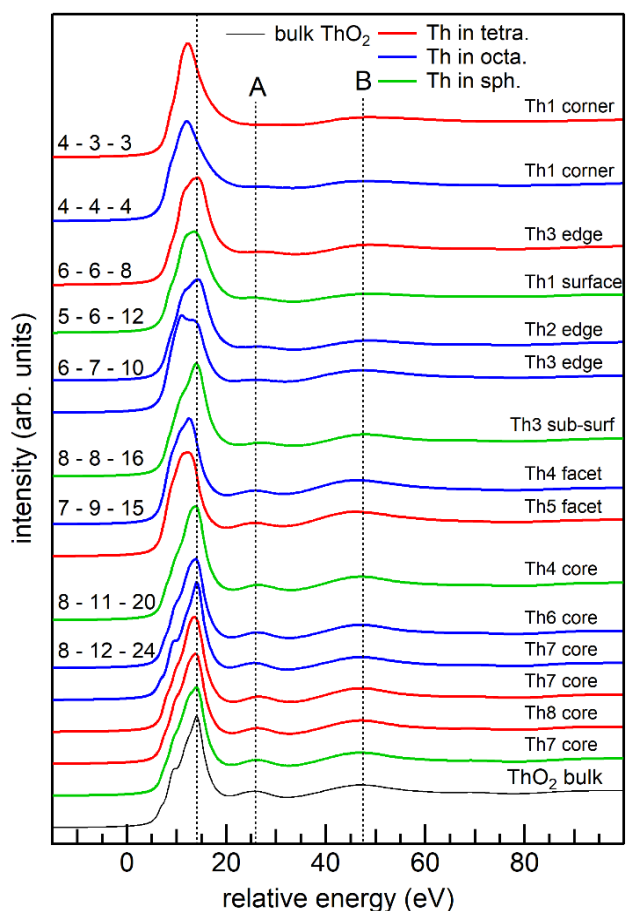

**Figure 4.** XANES simulations of single Th absorbers from the octahedral (blue lines), tetrahedral (red lines), and spherical (green lines) model NPs. A label with the reference Th# and the position it occupies in the NP is reported on the right of the spectrum. The XANES simulation of bulk $ThO_2$ is reported on the bottom of the graph (black thin line). The simulated spectra are grouped according to the coordination numbers of the first three coordination shells which are reported on the top-left for each group. Vertical lines to guide the eye have been added in correspondence of the white line, the post-edge feature A, and B.

Each model presents Th atoms having the coordination numbers of the first three coordination shells almost identical to that of bulk $ThO_2$, i.e., 8O – 12Th – 24O. Those Th can be assigned to the core of the NP. From Figure 3 we see that there are 10 core Th atoms in the tetrahedral NP (Th7 and Th8), 19 core Th in the octahedral NP (Th 5, Th6 and Th7), and 43 core Th in the spherical NP (Th5, Th6 and Th7). Th at the surface on the other hand have a reduced number of neighbouring atoms compared to Th in bulk $ThO_2$. The extreme case is that of a Th at the corner of a tetrahedral NP which has 4O – 3Th – 3O. The presence of sharp edges can be seen in the scaling of the coordination numbers of the first three shells, which in the case of the sphere increases more smoothly towards those of Th in bulk $ThO_2$ compared to the case of the octahedral and tetrahedral NPs. In our model NPs the surface Th atoms constitute 88%, 77% and 68% of the total cations for the tetrahedral, the octahedral and the spherical model, respectively. However, differentiate only between core and surface Th is not sufficient and inspection of the different kind of Th at the surface is important to understand their impact on the spectral shape.

To determine how the diverse local environments of Th atoms affect the XANES spectrum we performed atom-specific calculations with FDMNES. The XANES spectra of the non-equivalent Th absorbers are obtained with a single simulation run in which the relative shifts of the Fermi energy are calculated. Figure 4 shows the simulated spectra of selected non-equivalent Th absorbers for the three model NPs. The spectra are grouped according to the coordination numbers of the first three shells starting from the Th in bulk $ThO_2$ and scaling up to Th at the corners which have the lowest coordination numbers for the first three coordination shells. The spectrum of Th in bulk $ThO_2$ is reported on the bottom.. The post-edge feature A is particularly sensitive to the progressive decrease of neighbouring atoms and it practically disappears for coordinations lower than 7 – 9 – 15, i.e., for the more exposed cations at the surface. Feature B is also affected but at a lesser extent. On the other hand, Th atoms with coordination identical to Th in bulk $ThO_2$ are almost unchanged. Relevant spectral deviations from bulk $ThO_2$ are clearly observed at the WL. However, to correctly simulate the very first part of the XANES, a SCF loop to calculate the potential and the use of the FDM are more appropriate. At the same time, modifying the simulation approach in this direction needs a better description of the structure including local disorder and charge balance. We can therefore consider the remarkable spectral differences at the WL as an indication that this region of the spectrum may also be very sensitive to size effects and that improving the modelling of the structure at the surface would allow the use of a more sophisticated approach to simulations.

The simulation representing the XANES of the whole NP is given by the weighted average of all non-equivalent Th absorbers in the NP. Figure 5 shows the results obtained for the three model NPs considered. The experimental data on the biggest (33.8 nm) and smallest (2.5 nm) NPs are reported and compared with three groups of calculations, one for each model NP. Each group reports the simulation of bulk $ThO_2$ (black line) superimposed on the spectrum representing a specific model NP (red line). All three model NPs present spectral differences with bulk $ThO_2$. In the post-edge region the main effect is the lowering of feature A, in agreement with the experimental data. As expected, the spherical NP presents the smallest variations while the effect is more pronounced for the octahedral and tetrahedral models. A very slight broadening of feature B is also reproduced by the simulations. In general, the size effect observed on the post-edge of the experimental data is qualitatively reproduced by all three models and almost quantitatively by the octahedral and tetrahedral models. These results stem from the presence of Th absorbers with coordination numbers for the first coordination shells drastically reduced from the bulk values and from the specific collection of Th local environments which is directly determined by the size and the shape of the NP.

The flattening of feature A observed in the post edge region of Th $L_3$ edge HERFD XANES for very small NPs differs from the general broadening observed in XANES of nanosized materials. The latter is generally ascribed to the structural disorder associated with nanomaterials while what we observed is more specific and can conceal information on the structure of the nanoparticles. An ideal approach to understand the significance of small spectral differences in XANES is to obtain the atomic coordinates of the atoms composing the NP from state-of-the-art theoretical approaches and utilize them as structural input for *ab initio* codes designed for XANES.[4,10,16,17] This workflow to XANES and EXAFS analysis is becoming relatively common, with DFT and MD being the preferred methods to treat systems with a large number of atoms.[7,46,47] Despite the impressive improvements of computational resources and simulation methods, obtaining the structure of small actinide nanoparticles from first principles is still beyond the current possibilities[32] and a simpler approach to the analysis is necessary.

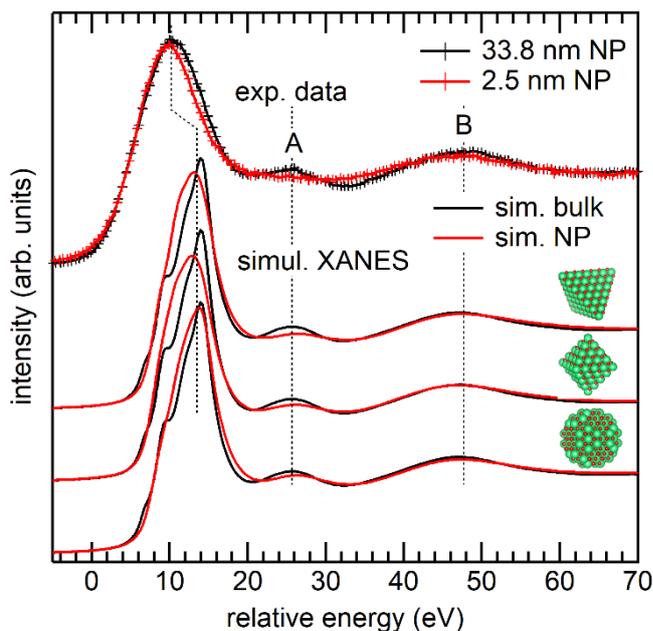

**Figure 5**. Experimental (top) and simulated (bottom) HERFD XANES at Th $L_3$ edge are compared. The data on big (33.8 nm, black curve) and small (2.5 nm, red curve) $ThO_2$ NPs are compared with the weighted average of all Th absorbers in the specific model NP (red lines). The spectrum of Th in bulk $ThO_2$ (black line) is reported for each model NP to highlight the spectral variations.

The precondition for a reliable analysis of XANES based on spectral simulations is the ability of the code of choice to reproduce correctly the reference spectra. In our case bulk $ThO_2$ is the only reference and as shown in Figure 2 the agreement between the less convoluted FDMNES simulation and data acquired with 0.5 eV resolution is excellent and the details of the spectral lineshape of HERFD XANES on $ThO_2$ bulk are well reproduced. Feature A is in the post edge region, which is known to be more sensitive to the local structure around the absorber rather than to its chemical state.

The model NPs that we built have all similar number of Th atoms, i.e., 84, 85 and 135 for the tetrahedral, octahedral and spherical NP, respectively. The average coordination numbers (CNs) of the first three coordination shells computed from the information resumed in Figure 3 are 5.95 – 8 – 13 for the tetrahedral, 6.59 – 8.47 – 14.12 for the octahedral and 6.87 –

8.89 – 17.07 for the spherical to be compared with 8 – 12 – 24 of the fluorite structure. The decrease of the average CNs quantifies the general impact of the surface atoms on the local structure of Th atoms. However, the consideration of the specific local environments of Th atoms present in each model provides additional information. The inspection of the spectra of single Th evidences how the progressive decrease of CNs is reflected in the post-edge feature A. In Figure 4, where spectra from single Th absorbers are grouped according to CNs independently from the NP model they belong to, the lowering of feature A is particularly pronounced for CNs lower than 7 – 9 – 15. For CNs between 8 – 12 – 24 and 7 – 9 – 15 the post-edge is only slightly changed, with both feature A and B undergoing a shift rather than a flattening or broadening. For lower CNs the changes are more important: feature A is flattened and features B is broadened. The spectra of single Th reveal the strong sensitivity of feature A to low CNs, i.e., to the more exposed Th at the surface. CNs lower than 7 – 9 – 15 correspond indeed to edges and corners in the tetrahedral and octahedral models and to Th at the surface in the spherical model.

From Figure 5, all the averages obtained by weighting the non-equivalent Th atoms spectra for each model NPs reproduce the trends observed on the experimental data. All models indeed, if compared to the simulation of bulk $ThO_2$ present a lowering of feature A and a broadening of feature B. However, the effect is very small for the spherical NP and more pronounced for the octahedral and the tetrahedral NPs. This reflects the percentage of Th with very low CNs for which the flattening of feature A is more pronounced as from Figure 4. The results presented illustrate how the different Th local environments resulting from the reduced size affect the XANES lineshape and what is the overall impact on the average spectrum. The size effect on the experimental spectra is stronger than in any of the models considered, indicating that additional factors impact on the intensity of feature A. The first to be considered is the local disorder, which in NP is expected to be stronger at the surface that in the bulk. Local distortions spread the ranges of Th – O and Th – Th distances and the contributions from nearest neighbours that sum up coherently in an unrelaxed structure undergo a de-phasing in a disorder structure. In this regards local disorder could induce additional decrease of feature A and bring the results to better match the experimental data. However, obtaining the extent and repartition of disorder in nanoparticles from first principles is not a trivial task,[4] especially for actinide-based nanomaterials. The interaction of the surface with the solvent should also be considered. Water molecules and hydroxyl groups attached to the surface will increase the number of O in the first coordination shell of surface Th. Inspection of the DOSs in Figure 2 indicates that the dominant contribution to feature A is Th d-DOS suggesting that few additional O from surfactants would induce minor effects.

The modelling approach we presented does not aim at determining the shape of the NPs from XANES. Instead, it provides an insight into the origin of size effects in XANES and demonstrate that a consistent contribution to the decrease of feature A is accounted by the more exposed Th atoms at the surface. We also suggest that local disorder at the surface is accounting for the additional lowering of feature A. We notice that the impact on XANES spectral shape of disorder in nanoparticles and clusters is poorly investigated and most of the studies using relaxed structures never discuss the impact that relaxation has on the results. Results presented here can guide future experimental investigations and suggest that a similar effect can be found in other oxide nanosystems. Post-edge features similar to feature A can be found in $L_3$ edge XANES of other actinide and lanthanide elements. The sensitivity of the first post-edge feature to the surface cations demonstrated in this work may be a point in common with other oxide NPs. To conclude, our work demonstrates that the post edge feature A of Th $L_3$ edge HERFD XANES spectrum is specifically sensitive to low coordinated surface cations. XANES and specifically feature A can therefore be sensitive to modifications of the surface atoms induced by chemical processes and surfactants. Some final remarks on the edge region and in particular on the WL can be drawn based on the data presented here. The edge region of the XANES is the most sensitive to the charge transfer, to the chemical state of the absorber and to the fine details of the local electronic structure. Previous studies have often observed that the size of NPs can affect the intensity of the WL and simple structural models like the one presented here have been used to reproduce the effect in the case of metal clusters.[48] A similar effect has been observed in actinide NPs and was ascribed to the reduced size rather than to charge transfer effects.[49,50] Our data do not present relevant variations of the WL intensity with decreasing size and our purely structural models do not predict strong effects on the WL intensity for small size $ThO_2$ NPs. We stress that the model NPs that we constructed are meant to represent only the structural effects introduced by the surface and not the electronic effects, therefore the spectral variations at the WL resulting from the simulations have to be taken with care. In Figure 4 the spectra of single Th absorbers show marked differences at the WL, hinting to the high sensitivity of this region of the spectrum. The differences at the WL on the average spectra are less pronounced but sizable. In particular, the three shoulders on the rising absorption edge that can be resolved if the resolution is increased (Figure 2) are smeared when size is reduced. These results are affected by strong approximations but suggest two important points when measuring HERFD XANES on actinide-based NPs: i) an increase in the energy resolution is fundamental to appreciate small spectral variations like the one expected when the size is reduced to the nanoscale; ii) a structural model of the surface accounting not only for the structural but also for the electronic effects is required as input for simulations that aims at an accurate description of the edge region.

## Conclusions

We presented HERFD XANES at Th $L_3$ edge on $ThO_2$ NPs of different sizes which show the flattening of the first post edge feature A for NPs with diameter < 2.5 nm. We performed XANES FDMNES simulations on three different structural models of small $ThO_2$ NPs considering the tetrahedral, the octahedral, and the spherical shape. Inspection of the simulations of single Th absorbers clearly shows that the progressive decrease of coordinating atoms induces a flattening of feature A, which is therefore particularly pronounced for the more exposed Th atoms at the surface. The comparison between experimental data and the simulations representing the model NPs shows that the size effect observed on the data is qualitatively reproduced by all the models and that the flattening of feature A is more pronounced and closer to the measured effect for models with higher number of the more exposed Th at the surface. The sensitivity of $ThO_2$ electronic structure to the number and the arrangement of surrounding atoms stems from the strong hybridization of Th d-DOS with both O and Th neighbours. In perspective, the sensitivity of the post-edge of Th $L_3$ edge HERFD XANES to the low coordinated atoms at the surface can be part of a toolkit of characterization techniques to investigate interactions at the surface and may not be limited to the case of $ThO_2$ but can be extended to other oxide nanosystems.

## Conflicts of interest

There are no conflicts to declare.

## Acknowledgements

The authors would like to acknowledge the ESRF for providing beamtime and C. Henriquet for providing technical support at ID20 beamline. This research was funded by European Research Council (ERC) under grant agreement No 759696. The synthesis of $ThO_2$ nanoparticles was funded by RFBR, according to the research projects No. 16-33-60043 mol_a_dk and No. 18-33-01067 mol_a. S.M.B. acknowledges support from the Swedish Research Council (research grant 2017-06465).

## References

1 H. Goesmann and C. Feldmann, *Angew. Chem. Int. Ed.*, 2010, **49**, 1362–1395.
2 S. J. L. Billinge and I. Levin, *Science*, 2007, **316**, 561–565.
3 G. Agostini, A. Piovano, L. Bertinetti, R. Pellegrini, G. Leofanti, E. Groppo and C. Lamberti, *J. Phys. Chem. C*, 2014, **118**, 4085–4094.
4 B. Gilbert, H. Zhang, F. Huang, J. F. Banfield, Y. Ren, D. Haskel, J. C. Lang, G. Srajer, A. Jürgensen and G. A. Waychunas, *J. Chem. Phys.*, 2004, **120**, 11785–11795.
5 B. Palosz, E. Grzanka, S. Gierlotka, S. Stel'makh, R. Pielaszek, U. Bismayer, J. Neuefeind, T. Proffen, R. V. Dreele and W. Palosz, *Z. Für Krist.*, 2002, **217**, 497–509.
6 F. Zhang, P. Wang, J. Koberstein, S. Khalid and S.-W. Chan, *Surf. Sci.*, 2004, **563**, 74–82.
7 A. Kuzmin and J. Chaboy, *IUCrJ*, 2014, **1**, 571–589.
8 A. I. Frenkel, *Chem. Soc. Rev.*, 2012, **41**, 8163.
9 L. Mino, G. Agostini, E. Borfecchia, D. Gianolio, A. Piovano, E. Gallo and C. Lamberti, *J. Phys. Appl. Phys.*, 2013, **46**, 423001.
10 A. L. Ankudinov, J. J. Rehr, J. J. Low and S. R. Bare, *J. Chem. Phys.*, 2002, **116**, 1911–1919.
11 F. Vila, J. J. Rehr, J. Kas, R. G. Nuzzo and A. I. Frenkel, *Phys. Rev. B*, 2008, **78**, 121404(R).
12 S. M. Butorin, K. O. Kvashnina, J. R. Vegelius, D. Meyer and D. K. Shuh, *Proc. Natl. Acad. Sci.*, 2016, **113**, 8093–8097.
13 K. O. Kvashnina and F. M. F. de Groot, *J. Electron Spectrosc. Relat. Phenom.*, 2014, **194**, 88–93.
14 P. Glatzel, T.-C. Weng, K. Kvashnina, J. Swarbrick, M. Sikora, E. Gallo, N. Smolentsev and R. A. Mori, *J. Electron Spectrosc. Relat. Phenom.*, 2013, **188**, 17–25.
15 J.-D. Cafun, K. O. Kvashnina, E. Casals, V. F. Puntes and P. Glatzel, *ACS Nano*, 2013, **7**, 10726–10732.
16 O. V. Safonova, A. A. Guda, C. Paun, N. Smolentsev, P. M. Abdala, G. Smolentsev, M. Nachtegaal, J. Szlachetko, M. A. Soldatov, A. V. Soldatov and J. A. van Bokhoven, *J. Phys. Chem. C*, 2014, **118**, 1974–1982.
17 A. Gorczyca, V. Moizan, C. Chizallet, O. Proux, W. Del Net, E. Lahera, J.-L. Hazemann, P. Raybaud and Y. Joly, *Angew. Chem.*, 2014, **126**, 12634–12637.
18 L. Amidani, A. Naldoni, M. Malvestuto, M. Marelli, P. Glatzel, V. Dal Santo and F. Boscherini, *Angew. Chem. Int. Ed.*, 2015, **54**, 5413–5416.
19 G. Rossi, M. Calizzi, L. Amidani, A. Migliori, F. Boscherini and L. Pasquini, *Phys. Rev. B*, 2017, **96**, 045303.
20 M. van der Linden, A. J. van Bunningen, L. Amidani, M. Bransen, H. Elnaggar, P. Glatzel, A. Meijerink and F. M. F. de Groot, *ACS Nano*, 2018, **12**, 12751–12760.
21 J. Timoshenko, A. Shivhare, R. W. J. Scott, D. Lu and A. I. Frenkel, *Phys. Chem. Chem. Phys.*, 2016, **18**, 19621–19630.
22 J. Timoshenko, D. Lu, Y. Lin and A. I. Frenkel, *J. Phys. Chem. Lett.*, 2017, **8**, 5091–5098.
23 J. Chaboy and S. Díaz-Moreno, *J. Phys. Chem. A*, 2011, **115**, 2345–2349.
24 S. N. Kalmykov and M. A. Denecke, Eds., *Actinide nanoparticle research*, Springer, Berlin ; London, 2010.
25 C. Walther and M. A. Denecke, *Chem. Rev.*, 2013, **113**, 995–1015.
26 H. Wu, Y. Yang and Y. C. Cao, *J. Am. Chem. Soc.*, 2006, **128**, 16522–16523.
27 O. N. Batuk, D. V. Szabó, M. A. Denecke, T. Vitova and S. N. Kalmykov, *Radiochim. Acta*, 2013, **101**, 233–240.
28 D. Hudry, J.-C. Griveau, C. Apostolidis, O. Walter, E. Colineau, G. Rasmussen, D. Wang, V. S. K. Chakravadhaluna, E. Courtois, C. Kübel and D. Meyer, *Nano Res.*, 2014, **7**, 119–131.
29 G. Wang, E. R. Batista and P. Yang, *Phys. Chem. Chem. Phys.*, 2018, **20**, 17563–17573.
30 D. Hudry, C. Apostolidis, O. Walter, T. Gouder, E. Courtois, C. Kübel and D. Meyer, *Chem. - Eur. J.*, 2013, **19**, 5297–5305.
31 D. Hudry, C. Apostolidis, O. Walter, A. Janßen, D. Manara, J.-C. Griveau, E. Colineau, T. Vitova, T. Prüßmann, D. Wang, C. Kübel and D. Meyer, *Chem. - Eur. J.*, 2014, **20**, 10431–10438.
32 B. Schimmelpfennig, in *Actinide Nanoparticle Research*, eds. S. N. Kalmykov and M. A. Denecke, Springer Berlin Heidelberg, Berlin, Heidelberg, 2011, pp. 187–193.
33 T. V. Plakhova, A. Y. Romanchuk, D. V. Lykhosherstova, A. E. Baranchikov, P. V. Dorovatovskii, R. D. Svetogorov, T. B. Shatalova, T. B. Egorova, A. L. Trigub, K. O. Kvashnina, V. K. Ivanov and S. N. Kalmykov, *Submitt. Dalton Trans.*
34 T. Reich, G. Bernhard, G. Geipel, H. Funke, C. Hennig, A. Roßberg, W. Matz, N. Schell and H. Nitsche, *Radiochim. Acta*, 2000, **88**, 633–637.
35 K. O. Kvashnina and A. C. Scheinost, *J. Synchrotron Radiat.*, 2016, **23**, 836–841.
36 M. Rovezzi, C. Lapras, A. Manceau, P. Glatzel and R. Verbeni, *Rev. Sci. Instrum.*, 2017, **88**, 013108.
37 M. Moretti Sala, K. Martel, C. Henriquet, A. Al Zein, L. Simonelli, C. J. Sahle, H. Gonzalez, M.-C. Lagier, C. Ponchut, S. Huotari, R. Verbeni, M. Krisch and G. Monaco, *J. Synchrotron Radiat.*, 2018, **25**, 580–591.
38 Y. Joly, O. Bunău, J. E. Lorenzo, R. M. Galéra, S. Grenier and B. Thompson, *J. Phys. Conf. Ser.*, 2009, **190**, 012007.

39 T. Vitova, K. O. Kvashnina, G. Nocton, G. Sukharina, M. A. Denecke, S. M. Butorin, M. Mazzanti, R. Caciuffo, A. Soldatov, T. Behrends and H. Geckeis, *Phys. Rev. B*, 2010, **82**, 235118.
40 K. O. Kvashnina, Y. O. Kvashnin, J. R. Vegelius, A. Bosak, P. M. Martin and S. M. Butorin, *Anal. Chem.*, 2015, **87**, 8772–8780.
41 H. Bao, P. Duan, J. Zhou, H. Cao, J. Li, H. Yu, Z. Jiang, H. Liu, L. Zhang, J. Lin, N. Chen, X. Lin, Y. Liu, Y. Huang and J.-Q. Wang, *Inorg. Chem.*, 2018, **57**, 11404–11413.
42 R. K. Behera and C. S. Deo, *J. Phys. Condens. Matter*, 2012, **24**, 215405.
43 N. V. Skorodumova, M. Baudin and K. Hermansson, *Phys. Rev. B*, 2004, **69**, 075401.
44 Z. L. Wang and X. Feng, *J. Phys. Chem. B*, 2003, **107**, 13563–13566.
45 A. Migani, K. M. Neyman and S. T. Bromley, *Chem. Commun.*, 2012, **48**, 4199.
46 S. T. Bromley, I. de P. R. Moreira, K. M. Neyman and F. Illas, *Chem. Soc. Rev.*, 2009, **38**, 2657.
47 A. R. Puigdollers, F. Illas and G. Pacchioni, *J. Phys. Chem. C*, 2016, **120**, 4392–4402.
48 D. Bazin, D. Sayers, J. J. Rehr and C. Mottet, *J. Phys. Chem. B*, 1997, **101**, 5332–5336.
49 J. Rothe, M. A. Denecke, V. Neck, R. Müller and J. I. Kim, *Inorg. Chem.*, 2002, **41**, 249–258.
50 J. Rothe, C. Walther, M. A. Denecke and T. Fanghänel, *Inorg. Chem.*, 2004, **43**, 4708–4718.